\documentclass[graybox]{svmult}

\usepackage{type1cm}        
\usepackage{makeidx}         
\usepackage{graphicx}        
\usepackage{multicol}        
\usepackage[bottom]{footmisc}

\usepackage{url}%
\usepackage{color} %
\usepackage{newtxtext}       %
\usepackage[varvw]{newtxmath}       
\usepackage[T1]{fontenc}

\makeindex             

\begin{document}

\title*{RHIC Program, Its Origin and Early Results}

\author{Wit Busza, John W. Harris, and Shoji Nagamiya }

\institute{Wit Busza  \at    MIT;  \email{busza@mit.edu}
\and
John W. Harris  \at    Yale University;  \email{john.harris@yale.edu}
\and 
Shoji Nagamiya  \at    KEK;  \email{nagamiya@post.kek.jp}}

%
%
\maketitle

\abstract{At the Brookhaven National Laboratory, experimental efforts with heavy-ion accelerators started at the AGS synchrotron in 1984 and then at the Relativistic Heavy Ion Collider (RHIC) in 1991. This chapter describes how several scientific collaborations were established and how features of the quark-gluon plasma were revealed from the four RHIC experiments during the first five years of RHIC operation.}

\section{RHIC Origins}
\label{RHIC_origins}

By the late 1970s an exponentially growing community was born, interested in studying ultra-relativistic heavy-ion collisions.\footnote{See previous chapter in this book by W. Busza and W. Zajc} It was a merger of particle physicists interested in the mechanism of multiparticle production in high energy hadronic collisions and nuclear physicists interested in investigating the nuclear equation of state and what happens when nuclear matter is compressed. These interests were further heightened by the discovery of quarks and gluons, asymptotic freedom, the existence of neutron stars, and the development of the theory of Quantum Chromo Dynamics (QCD). This led to the realization that, in essence, the two groups were interested in the same physics; the condensed matter of QCD at ultra-high energy density and temperature. Furthermore, it was realized that this field had some fascinating and important questions that needed to be answered. Is the vacuum around us unique? Do different forms of nuclear matter exist? Does a phase of QCD matter exist in which quarks and gluons are free? Does there exist a phase in which chiral symmetry is restored? What does the complete phase diagram of QCD matter look like? Can any of these phases be created and studied in the laboratory? An added interest in these questions was the realization that in the early universe, about 10 microseconds after the big bang, the hadronic part of the entire universe was thought to be in a gas-like state of free and noninteracting quarks and gluons, a state to which the name ``quark-gluon plasma” was given, or ``QGP”\footnote{Prior to RHIC, the community had adopted the term ``QGP” to describe any system that is best described in terms of quark and gluon degrees of freedom. There was already some caution from theory that the QGP could not be an ideal gas of deconfined, weakly interacting quarks and gluons. As soon as the RHIC results exhibited strong collectivity, it was realized that the QGP had to be strongly interacting.} for short.

Interest in these questions led to numerous workshops and conferences. It also stimulated discussions on whether current accelerators could be used to address some of these issues and
whether new accelerators or colliders needed to be constructed. 
After a landmark recommendation\footnote{The US Nuclear Science Advisory Committee identified the scientific opportunities of a relativistic heavy-ion collider and recommended it as the next major construction project for nuclear science in the US. See {https://science.osti.gov/-/media/np/nsac/pdf/docs/lrp\_1983.pdf}} in 1983 and the unanticipated availability of an unfinished ISABELLE/CBA collider
complex at Brookhaven National Laboratory (BNL) \cite{Samios:2007zz}, there was considerable interest in a possible relativistic heavy-ion collider at BNL.
A special round-table discussion by a distinguished international panel of nuclear and high-energy physicists was held at the 1983 Quark Matter Conference at BNL to discuss the future of relativistic heavy-ion physics, and led to the RHIC
experimental program followed by an LHC heavy-ion program at CERN
 \cite{Ludlam:1984klg}.



An obvious concern, which ultimately could only be answered experimentally, was whether any existing accelerator, or the future RHIC, could produce a state that has sufficiently high energy density and lasts for a sufficiently long time to create an interesting new kind of QCD matter, such as the QGP.
In the 1970s/1980s, based on the assumption that a QGP was a gas of noninteracting quarks and gluons, knowledge of the masses and sizes of hadrons, and the Hagedorn temperature \cite{Hagedorn} (where there is a very rapid increase in the number of hadron levels with mass), it was believed that for the creation of a QGP one needed an energy density of the order of or greater than 1 GeV/fm$^3$.

\begin{figure}[h]
\begin{center}
\includegraphics[scale=0.4]{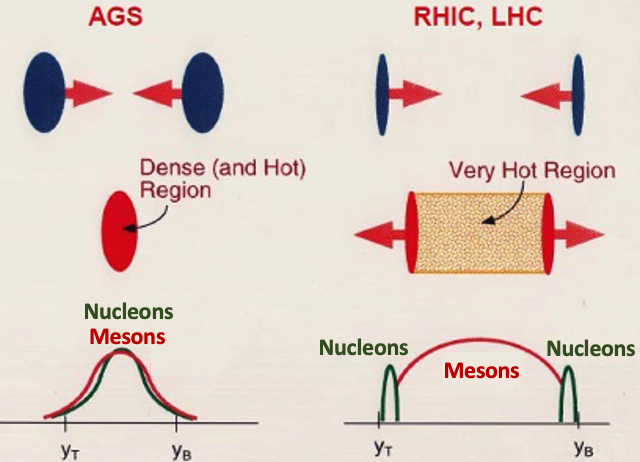}
\end{center}
\caption{Schematic diagram of a collision of two nuclei at the AGS (left column), and at RHIC and the LHC (right column), as described in the text. At bottom are shown the corresponding rapidity (y) distributions after the collision.}
\label{figures:AGS_RHIC_LHC} 
\end{figure}

The community was full of optimism that 1 GeV/fm$^3$ would be achieved at RHIC, and further encouraged by extrapolations of Fermilab p-A data \cite{Nuclear_stopping}, which showed that high-energy protons that traverse large nuclei lose about 2 units of rapidity. Thus, suggesting that in head-on heavy-ion collisions about 85\% of the incident energy will be deposited in the collision. Experiments confirmed these estimates \cite{Nagamiya:1988ge}. At the AGS energy of 14 A-GeV for a projectile impinging on a target at rest, it was observed that the two nuclei stop each other in their center of mass, while on the other hand they penetrate through each other at the higher SPS energies of 200 A-GeV. Thus, it was expected that at the much higher energies of RHIC, the colliding nuclei would penetrate each other completely and create a hot baryon-free region, as schematically illustrated in Fig.~\ref{figures:AGS_RHIC_LHC}. Although expectations were that at RHIC energies the all-important energy density in the hot region would exceed 1 GeV/fm$^3$, experimental confirmation was necessary. In short, it needed the construction of RHIC and its detectors! 

Given the above interest and expectations, a much-publicized workshop was held at BNL on July 4, 1990, to initiate discussion of the RHIC experiments. It was a particularly important workshop in that the BNL RHIC project had just been included in the President's proposed FY 1991 budget as a new construction item, and a call for letters of intent (LoIs) for RHIC experiments had been issued at the BNL Users meeting in May 1990. A workshop was held in July 1990 to form collaborations and prepare LoIs. Eight LoIs\footnote{RLOI-2,  TALES/SPARHC: Two-Arm Electron/Photon/Hadron Spectrometer; RLOI-3,  Search for a Quark-Gluon Plasma and Other New Phenomena with a $4\pi$ Tracking TPC Magnetic Spectrometer at RHIC; RLOI-4,  A Dimuon Spectrometer for RHIC Measurements of Muon Pairs, Vector Mesons, and Photons; RLOI-5,  STAR: An Experiment on Particle and Jet Production at Midrapidity; RLOI-6,  MARS (later became PHOBOS): A Modular Array for RHIC Spectra; RLOI-7,  OASIS: Open Axially Symmetric Ion Spectrometer; RLOI-8,  (later became BRAHMS) Forward Angle Hadron Spectrometer Experiment at RHIC; RLOI-9, High p$_T$ Photons, Charged Particles and Jets at RHIC.} to study A-A collisions at RHIC were submitted to the BNL High Energy Nuclear Physics Program Advisory Committee in November 1990. Guided by almost a year of discussions, including a community-wide meeting in April 1991, four proposals for experiments were submitted in 1991: STAR to measure charged particles over 4$\pi$; OASIS to measure identified hadrons, photons, and $e^+e^-$ pairs; Dimuon to measure $\mu^+\mu^-$ pairs; and TALES/SPARHC to measure $e^+e^-$ pairs. The proposed detectors, among them, covered the full phase space of all of the produced particles and would be more than adequate to investigate heavy-ion collisions at RHIC. However, obviously all were not within the budget of the RHIC program, and thus decisions had to be made on what to build. These proposals were sent to BNL and a meeting of the BNL High Energy Nuclear Physics Program Advisory Committee was held in September 1991 to consider the proposals. 


Management decided that there would be two large general-purpose detectors, but with different physics strengths.
The STAR experiment was approved, while the three others were advised to merge and then formed the PHENIX experiment in 1992. STAR, PHENIX and a smaller experiment PHOBOS were given final approval at the BNL High Energy Nuclear Physics Program Advisory Committee meeting in September 1992.

\begin{figure}[ht]
\begin{center}
\includegraphics[scale=0.3]{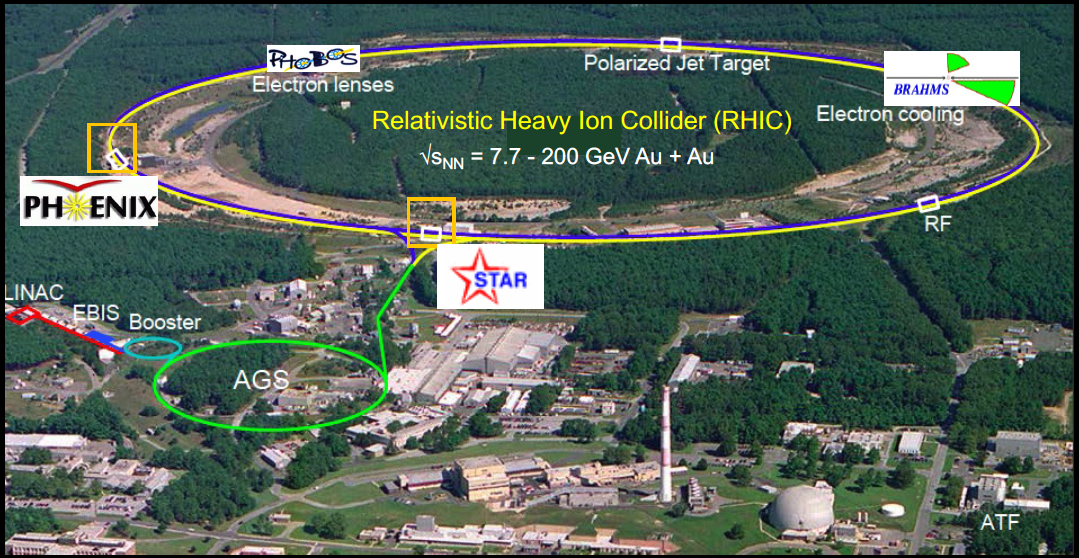}
\end{center}
\caption{The RHIC Complex and its four detector experiments. See \cite{RHIC_NIM} for reference and details. }
\label{figures:RHIC}       
\end{figure}

The funding from the U.S. Department of Energy was set at \$30M for each of the large detectors, and additional resources, if needed, would have to be arranged by the individual experimental groups. BRAHMS, a two-arm hadron spectrometer, was approved later as one of the small detectors. The RHIC complex with its four experiments is shown in Fig.~\ref{figures:RHIC} and described in \cite{RHIC_NIM}.

\section{The STAR Experiment at RHIC}
\label{STAR_Experiment}

\subsection{Conception of the STAR Experiment at RHIC}
\label{STAR_Conception}


The concept and design of the STAR experiment were created after an almost-year-long series of weekly meetings and discussions. A series of
RHIC Planning Meetings was held at LBL (October '89 - July '90) where theorists and experimentalists were invited to discuss potential defining measurements at a future Relativistic Heavy Ion Collider (RHIC). In total 24 presentations were made on various topics on detector technologies and relativistic heavy-ion theory over that time frame, with the goal of deciding on the specifications of an experiment that would address important  questions of the field.


In addition, it was recognized that a new large-scale RHIC experiment would be novel to nuclear physics and \textit{new detector techniques} would need to be considered. Those included time projection chambers (TPCs), ring-imaging Cherenkov (RICH) detectors, photon detectors, charge-coupled devices (CCDs), smart calorimeters, scintillating fibers, and more. Furthermore, \textit{data acquisition} systems would be encumbered by large event sizes, and the continued development of high-density fast integrated electronics would further exacerbate the problem. Thus, rapid online data reduction would be necessary, but advances in large-scale fast data storage devices and media would still be needed. Together, several complex detector systems would be integrated into one experiment.

There were \textit{sociological and operational factors} that had to be considered. Although perhaps not new to high-energy physicists, this was new to our community. A large research collaboration would need to form in order to carry out all aspects of such an experiment - including its design, integration of detector systems, construction, operation, and physics. This would require better communication, planning, organization, and management. Lessons would be learned, and advice from the high-energy community was very welcome. The long-term aspects of such an experiment also required a new look at traditional hiring practices in experimental nuclear physics. We were getting prepared for the challenge, but was nuclear physics in general prepared for RHIC? 

\subsection{STAR Physics and Experiment}
\label{STAR}
The physics goals and a conceptual design were starting to converge by January 1990. The primary focus for RHIC and the STAR physics program in general was the study of QCD
at high energy density
 and temperature. The STAR approach was to have a large acceptance detector that would maximize
the information recorded per collision.  
In the absence of definitive signatures for the QGP, the high track multiplicities should allow the extraction of global observables such as centrality, temperature, reaction plane, and the mean transverse energy〈E$_T$〉. 

In a search for possible QGP signatures, the STAR experiment was initially focused on the measurement and correlation of as many observables as possible on an event-by-event basis. 
The physics goals could be divided into i) the study of soft (non-perturbative) physics processes, i.e., low p$_T$ hadron production, and ii) the study of hard (perturbative) QCD processes, i.e. jet, mini-jet, and high-p$_T$ particle production. The physics program also included the measurement of the spin structure function of the proton, and the study of photon and pomeron interactions from the large electromagnetic fields created by the passing heavy ions at RHIC. 

\begin{figure}[ht]
\begin{center}
\includegraphics[scale=0.35]{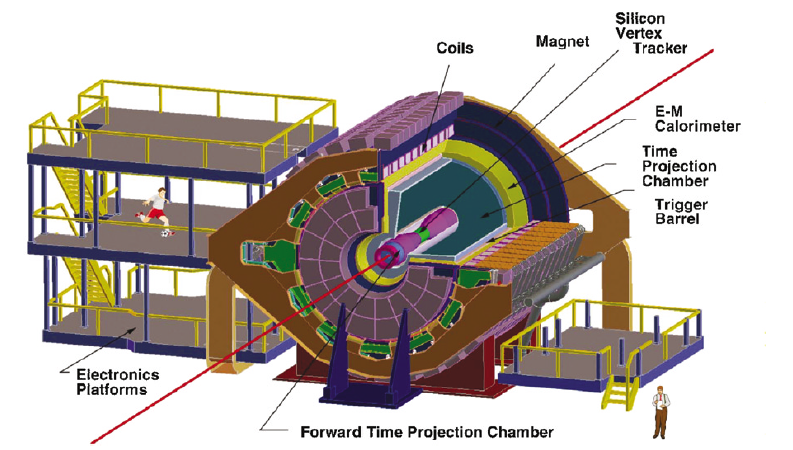}
\end{center}
\caption{The STAR Detector and its sub-detector systems as labeled. See \cite{STAR_NIM} and references therein for details.}
\label{figures:STAR_detector}       
\end{figure}

To meet these goals, STAR was designed to have large uniform acceptance, good momentum and two-track resolution, and particle identification capabilities to facilitate measurement of many observables. The STAR detector is shown in Fig.~\ref{figures:STAR_detector} and described in \cite{STAR_NIM}.
Examples of such measurements included particle spectra, flavor composition, particle interferometry, as well as density fluctuations of energy, entropy, and multiplicity in azimuth and pseudorapidity. The remnants of hard-scattered partons were used as a penetrating probe of the QGP, and expected to provide important new information on the nucleon structure function and parton shadowing in nuclei. 

An initial physics goal of STAR was to measure the relative concentration of strange and non-strange quarks on an event-ensemble basis, as well as strange and anti-strange baryons over a wide rapidity interval about mid-rapidity. Enhancements in strange antibaryon content were predicted in QGP formation, with multi-strange baryons expected to be even more sensitive to the existence of the QGP. \cite{Muller}

Measurements of strange and anti-strange baryons required detection of secondary decay vertices of particles within a distance c$\tau \sim$ 2-8 cm from the primary collision vertex. The tracking capabilities of a time projection chamber in STAR provided an azimuthally complete acceptance over a large rapidity interval. These qualities were essential for measuring the production of strangeness and open charm in collisions at RHIC in the search for the QGP. The STAR experiment, with its multiparticle and vertex tracking capabilities, was able to achieve these tracking goals. 

An important aspect of the STAR experiment was the ability to measure and trigger on events with large charged-particle multiplicities, total electromagnetic energy, high-p$_T$ $\pi^o$s, jet energies and single and multiple high-p$_T$ particles (including $\pi^o$s).  STAR measured two-particle correlations over a large acceptance at mid-rapidity with good momentum resolution and using Hanbury-Brown and Twiss (HBT) interferometry techniques \cite{HBT} was able to determine the size and lifetime of the particle-emitting source. It was anticipated that measurements would also be made in pp collisions and a range of p-A and A-A collisions.

\begin{figure}[ht]
\begin{center}
\includegraphics[scale=0.4]{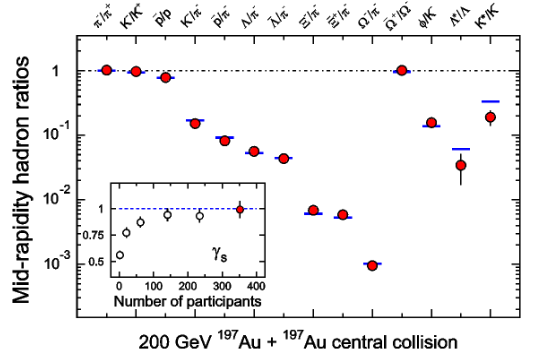}
\end{center}
\caption{Ratios of integrated mid-rapidity yields for various  identified hadrons in central
Au-Au collisions at $\sqrt{s_{NN}}$ = 200 GeV measured in STAR \cite{STAR:2005gfr}. Statistical model fits (horizontal bars) to the measured
ratios give fit parameters T$_{ch}$ = 163 $\pm$ 4 MeV, $\mu_B$ = 24 $\pm$ 4 MeV, $\gamma_s$ = 0.99 $\pm$ 0.07. The variation of the strangeness saturation parameter $\gamma_s$ with centrality is shown in the inset. See text.}
\label{figures:STAR_hadron_ratios_thermal}       
\end{figure}

A key early result from STAR was the measurement of the spectra and integrated yield ratios of identified hadrons at mid-rapidity in central Au-Au collisions. See \cite{STAR:2005gfr} for details. 
Fig.~\ref{figures:STAR_hadron_ratios_thermal}
shows these ratios along with a fit from the statistical model. The yields of multi-strange baryons $\Xi$ and $\Omega$ were observed to be considerably enhanced in central Au-Au collisions compared to pp collisions at similar energies. 
The measured ratios constrain
the values of the system temperature and the baryon chemical potential $\mu_B$ at chemical freeze-out T$_{ch}$.  Furthermore, the strangeness saturation parameter $\gamma_s$ increases with centrality as seen in the inset, where for the most cental collisions 
 $\gamma_s$ = 0.99 $\pm$ 0.07 compared to the much lower value for pp collisions, which is the leftmost point of the inset. 
The fits obtained from the ratios using the formulation of \cite{Thermal_fits}
are consistent with the assumption that the system is in thermal and chemical equilibrium at 
T$_{ch}$= 163 $\pm$ 5 MeV.


\begin{figure}[ht]
\begin{center}
\includegraphics[scale=0.45]{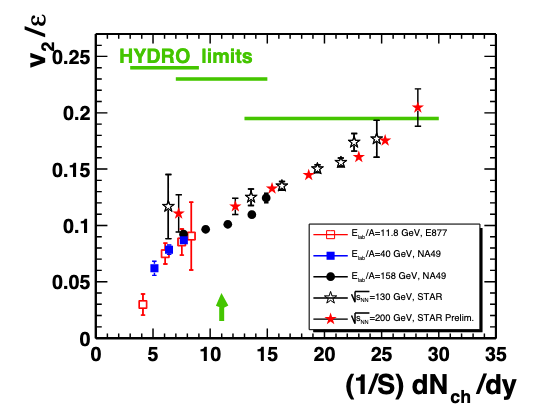}
\end{center}
\caption{Elliptic flow
measurements at various energies and centralities showing v$_2/\epsilon$ (elliptic flow/initial
spatial eccentricity) as a function of charged-particle rapidity density per unit transverse overlap area in the A-A collision. 
The v$_2/\epsilon$ rises smoothly versus energy and centrality. The highest energy STAR results \cite{STAR:2000ekf} are consistent with limiting hydrodynamic expectations for an ideal relativistic fluid represented by green horizontal bars drawn for AGS, SPS and RHIC energies. From \cite{NA49_hydro} .}
\label{figures:STAR_v2_hydro_limit}       
\end{figure}

A significant result from RHIC was the measurement of collective flow in Au-Au collisions \cite{STAR:2000ekf}. The first indication that the hydrodynamic limit was reached at RHIC at the highest multiplicity densities can be seen in Fig.~\ref{figures:STAR_v2_hydro_limit}. The rightmost points represent the highest multiplicities  in near-central collisions in STAR. There, the observed ratio of elliptic flow / shear viscosity (v$_2/\epsilon$) is consistent with rapid thermalization (within $\sim$ 1 fm / c) and reaches the limiting hydrodynamic expectations for an ideal relativistic fluid \cite{STAR:2005gfr}. The hydrodynamic limits are represented by green horizontal bars drawn for each range of collision energies
for a particular choice of equation of state that
assumes no phase transition in the system that is produced. See 
\cite{NA49_hydro} for details. 
This means that the behavior of the system at RHIC can be described by hydrodynamics, whereas at the lower energies, the system does not reach the hydrodynamic limit. 

Once it was established that there was a highly-excited thermal medium, and collective flow at the highest energies in central A-A collisions, the experiments sought to find out whether high-momentum particles lose energy as they traverse the excited QGP medium. In order to compare with normal nuclear matter, an observable called the nuclear modification factor R$_{AA}$ was measured and given by 

\begin{equation}
R_{AA} =  N_{AA}\hspace{1mm} /\hspace{1mm} <T_{AA}> \times \hspace{1mm}\sigma_{pp}
\end{equation}

\hspace{-5mm} where $N_{AA}$ is the number of observed particles per event of interest in A-A collisions, $<T_{AA}>$ is the average value of the nuclear thickness function, and $\sigma_{pp}$ is the
particle production cross section in pp collisions at the same collision energy.

\begin{figure}[ht]
\begin{center}
\includegraphics[scale=0.38]{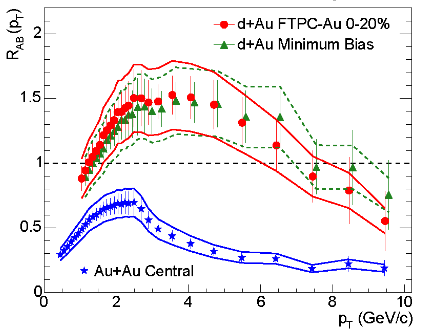}
\includegraphics[scale=0.38]{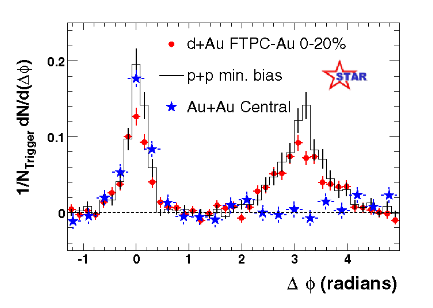}
\end{center}
\caption{(Left panel) R$_{AA}$ for central Au-Au collisions and R$_{AB}$ for the asymmetric d-Au collisions at $\sqrt{s_{NN}}$ = 200 GeV as measured by STAR \cite{STAR:2003pjh}. 
(Right panel) Dihadron azimuthal correlations at high p$_T$ for p + p, central d + Au and central Au + Au collisions (with background subtracted) at $\sqrt{s_{NN}}$ = 200 GeV from STAR \cite{STAR:2003pjh,STAR:2002svs}.}
\label{figures:STAR_jets}       
\end{figure}

 Fig.~\ref{figures:STAR_jets} (left panel) is the nuclear modification factor measured for Au-Au central collisions and d-Au collisions as a function of p$_T$ \cite{STAR:2003pjh}. The central Au-Au collisions exhibit a suppression of particles at high p$_T$ compared to the d-Au collisions, indicating that the particles lose energy as they traverse the QGP medium formed in the central Au-Au collisions. Fig.~\ref{figures:STAR_jets} (right panel) shows angular correlations of particles at azimuthal angles $\Delta\phi$ relative to the peak ($\Delta\phi = 0$) in pp, d-Au, and Au-Au collisions \cite{STAR:2002svs}. The peak represents the jet cone of particles around the high-momentum trigger particle. The peak of particles at 180$^\circ$ ($\Delta\phi = \pi$) away from the trigger particle in pp and d-Au collisions is due to binary scattering of incident particles.  However, high-momentum particles in central Au-Au collisions have no peak at 180$^\circ$ ($\Delta\phi = \pi$) away from the trigger, which indicates quenching of the jet that traversed the highly excited QGP in the direction opposite the trigger.

After the first years of RHIC operation and physics, STAR proposed various new detectors and upgrades that have since become an integral part of STAR and its physics. STAR added barrel and end-cap time-of-flight detectors, a compact muon detector, intermediate and forward tracking detectors, a forward meson spectrometer, an event-plane detector, a higher-rate data acquisition system, faster TPC readout and extended coverage of the TPC, and a forward silicon tracking and calorimeter upgrade. 

These new detector upgrades enabled a broad expansion of the STAR heavy-ion physics program that has included interesting results on the suppression of quarkonia \cite{STAR:2022rpk}, thermal dileptons \cite{STAR:2023wta}, ultra-peripheral collisions \cite{STAR:2023nos}, an unexpected chiral magnetic effect \cite{STAR:2023gzg}, and new inquiry into how baryon charge is carried in these collisions \cite{STAR:2024lvy}.
Additional investigation has continued into the behavior of open heavy-flavor particles,  higher-order flow harmonics and fluctuations, correlation lengths and susceptibilities of the QGP, jet quenching and substructure, and links to the nuclear equation of state via beam energy scans down below SPS energies.\footnote{https://drupal.star.bnl.gov/STAR/files/BES{\_}WPII{\_}ver6.9{\_}Cover.pdf}
STAR has discovered new light antimatter nuclei and antimatter hypernuclei, for example ${}_{\bar{{\Lambda }}}{}^{{4}}\bar{{H}}$; see \cite{STAR:2023fbc} and the references therein for the achievements of the STAR antimatter nucleus program. In addition, recent observation by STAR of an unexpectedly large global spin alignment of the vector mesons (spin 1) $\phi$ and K$^{*o}$ appears to only be describable in terms of strong force fields in QCD.
\cite{STAR:2022fan}. The detector performance and physics of the above studies continue to be presented at semiannual Quark Matter meetings.\footnote{See https://inspirehep.net/conferences?sort=dateasc\&size=25\&page=1\&start\_date=all\&\&q=QM} 

%
%

\section{The PHENIX Experiment at RHIC}
\label{PHENIX}

\begin{figure}[ht]
\begin{center}
\includegraphics[scale=1.2]{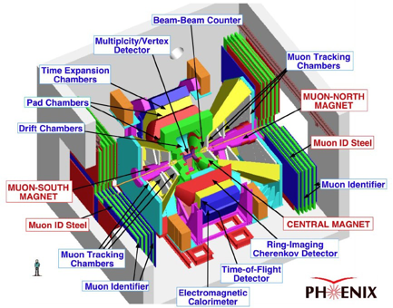}
\end{center}
\caption{The PHENIX Detector and subsystems as labeled. See \cite{PHENIX_NIM} for reference and details.}
\label{figures:PHENIX}
\end{figure}

Immediately after the PHENIX collaboration was formed, the PHENIX goals and experiment were proposed in June 1992. The design proposed at that time is shown in Fig.~\ref{figures:PHENIX}. The US-Japan High Energy program sponsored by KEK held a review and decided to add $\geq$ \$10M to support the construction of a time-of-flight detector to measure hadrons, a ring imaging detector, and a transition radiation detector to measure $e^+e^-$ pairs over a limited solid angle. In addition, an electromagnetic (EM) calorimeter was included to measure photons. A forward arm of PHENIX was planned for the measurement of $\mu^+\mu^-$ pairs, although resources were still needed to cover this portion. 

In 1993, the RIKEN group proposed a polarized proton spin physics program at RHIC with Siberian snakes and spin rotators for both PHENIX and STAR. The RIKEN-Spin group joined PHENIX and the polarized proton program with one muon arm for PHENIX was approved by RIKEN in 1994. A second muon arm was approved by the US DOE in 1995, leading to the final configuration of PHENIX \cite{PHENIX_NIM}, which includes the two muon arms (RIKEN and DOE) shown in Fig.~\ref{figures:PHENIX}.


\begin{figure}[ht]
\begin{center}
\includegraphics[scale=1.]{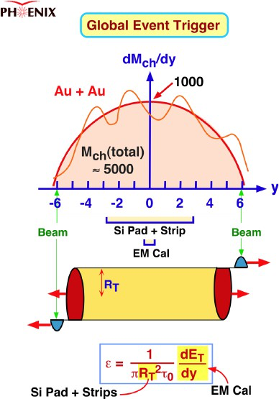}
\end{center}
\caption{PHENIX diagrams that illustrate its multiplicity measurement M$_{ch}$(top) that is used to deduce R$_T$ (middle) with its calorimeter measurements of dE$_T$/dy and indicating (bottom) how they are used to determine the energy density $\epsilon$. From \cite{Nagamiya:2015raa}.}
\label{figures:Event_trigger}
\end{figure}

The initial physics goals of PHENIX were expected to be achieved in three steps, as illustrated in Fig.~\ref{figures:Event_trigger} 
\cite{Nagamiya:2015raa}. First, PHENIX would measure the multiplicity up to the highest multiplicity, as seen in the top of the figure. Next, it would measure dE$_T$/dy with the EM calorimeters to determine R$_T$ as illustrated in the middle of Fig.~\ref{figures:Event_trigger}.  Since the energy density $\epsilon$ can be given by

\begin{equation}
\epsilon = E/V = (dE_T/dy) \hspace{1mm} / \hspace{1mm} (\pi R^2_T c\tau_o)
\end{equation}

\hspace{-5mm} the third step was to measure as many potential signals as possible as a function of dE$_T$/dy (or equivalently as a function of $\epsilon$).




%

It was predicted that the c$\bar{c}$ bound state would no longer exist at sufficiently high temperatures due to the Debye screening of QCD interactions between the c and $\bar{c}$ quarks. Since J/$\psi$ is a bound state of c and $\bar{c}$, J/$\psi$ particle production would be strongly suppressed \cite{Matsui-Satz}. As the radius of $\psi$' is larger than J/$\psi$, the degree of suppression would be stronger for $\psi$' than for J/$\psi$. PHENIX planned to explore the systematic measurement of these vector mesons by detection of dielectrons and dimuons.
In the 1990s, other predictions existed, such as measurements of QGP radiation, since the photon production mechanism differs considerably between a hadron gas and a quark-gluon gas \cite{Ruuskanen}.
Unfortunately, most of the predicted signals in the 1990s were not all easily accessible to experiment, justifying the experimental approach to measure multiple signals. 

\begin{figure}[ht]
\begin{center}
\includegraphics[scale=0.35]{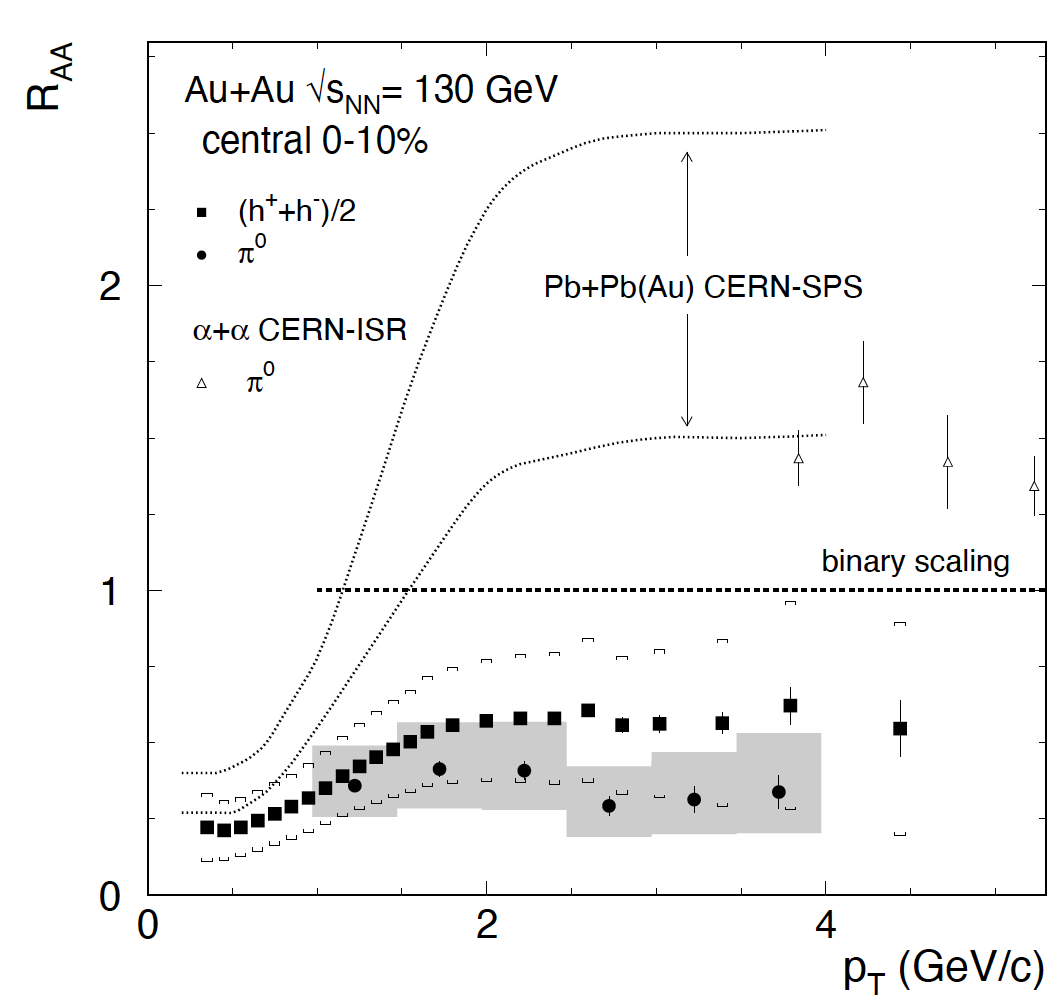}
\end{center}
\caption{Jet quenching measurements by PHENIX at the early RHIC energy of $\sqrt{s_{NN}}$ =130 GeV. Also shown are results from Pb-Pb measurements at the CERN-SPS ($\sqrt{s_{NN}}$ = 17 GeV) and $\alpha\alpha$ measurements at the CERN-ISR ($\sqrt{s_{NN}}$ = 31 GeV). From \cite{PHENIX_RAA}. See text.}
\label{figures:RHIC_ISR_RAA}
\end{figure}

In the early running of RHIC, while PHENIX was accumulating the statistics necessary to be able to address the prediction of suppression of J /$\psi$, it also collected data that significantly contributed to our understanding of jet quenching \cite{PHENIX-NIM}. Fig.~\ref{figures:RHIC_ISR_RAA} shows the early charged-hadron and $\pi^o$ measurements from PHENIX in central Au-Au collisions at $\sqrt{s_{NN}}$ = 130 GeV. The suppression observed in these measurements is compared with previous Pb-Pb measurements at CERN-SPS and $\alpha\alpha$ measurements at the CERN-ISR at $\sqrt{s_{NN}}$ = 17 and 31 GeV, respectively. The lower energy measurements exhibit an enhancement that has been attributed to an anomalous nuclear enhancement, called the Cronin effect \cite{Cronin}. See \cite{PHENIX_RAA} for more details.

\begin{figure}[ht]
\begin{center}
\includegraphics[scale=0.65]{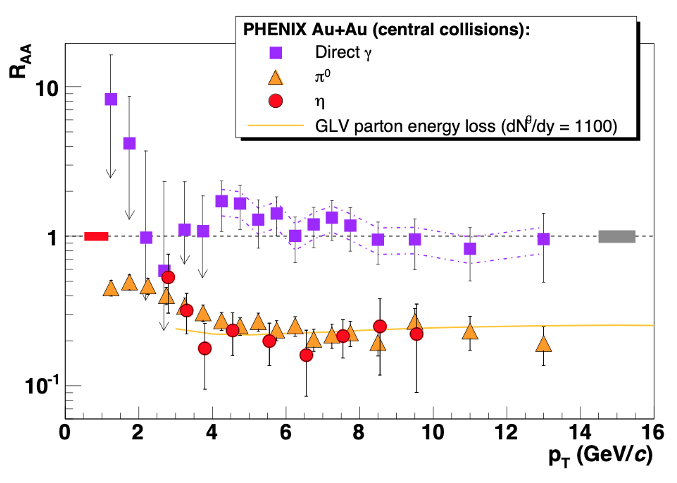}
\end{center}
\caption{R$_{AA}$ measured by PHENIX for $\pi^o$, $\eta$, and direct
$\gamma$ as a function of
p$_T$ in central $\sqrt{s_{NN}}$ = 200 GeV Au-Au collisions. See text and \cite{PHENIX_RAA} for details.}
\label{figures:PHENIX_gamma_RAA}
\end{figure}

Shown in Fig.~\ref{figures:PHENIX_gamma_RAA} 
are PHENIX $R_{AA}$ measurements of identified $\pi^o$, $\eta$ and direct $\gamma$ out to large values of p$_T$. The $\pi^o$ and $\eta$ are suppressed, while the $\gamma$ is not. Since $\gamma$ does
not
interact via strong
interactions, the fact that $R_{AA}(\gamma) \simeq 1$
provides strong evidence that the observed suppression of hadrons ($\pi^o$ and $\eta$) is due to their strong final-state interaction as they traverse the QGP.

\begin{figure}[ht]
\begin{center}
\includegraphics[scale=0.3]{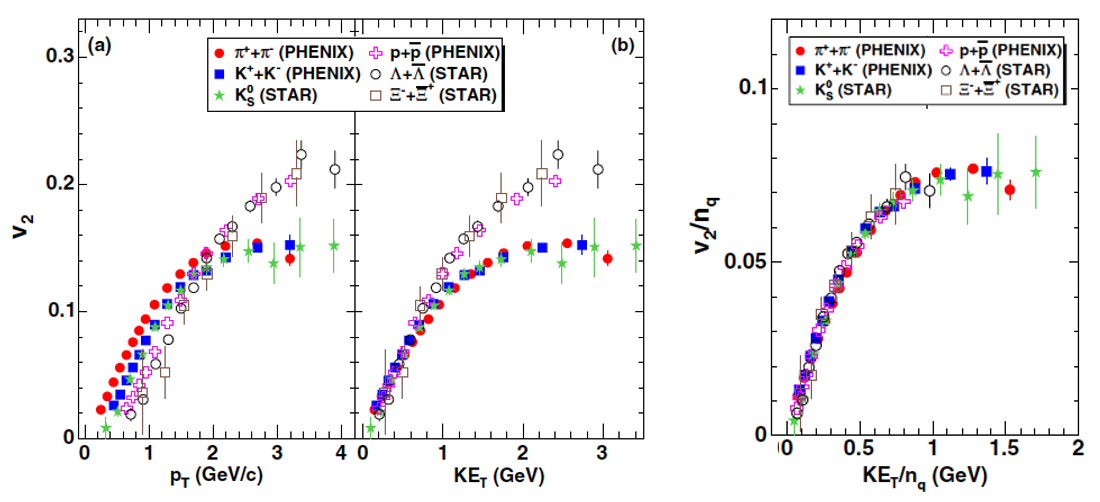}
\end{center}
\caption{Left double panel: a) Elliptic flow (v$_2$) for particles identified in the inset by PHENIX and STAR versus particle p$_T$. b) Elliptic flow versus particle transverse kinetic energy KE$_T$. Right panel: Elliptic flow per quark v$_2$/n$_q$ versus the transverse kinetic energy per quark KE$_T$/n$_q$, revealing  that constituent quark number scaling holds, which is observed by all particles falling on the same curve. From \cite{PHENIX:2006dpn}.}
\label{figures:v2_nq}
\end{figure}

PHENIX also collected important results on the elliptic flow that STAR observed. The PHENIX and STAR elliptic flow v$_2$ for identified particles are presented in Fig.~\ref{figures:v2_nq} as a function of a) particle p$_T$ and b) particle transverse kinetic energy KE$_T$, where KE$_T$ = m$_T$ - m, taken from 
\cite{PHENIX:2006dpn}. The data in the left panels appear to separate into two curves, one for the mesons and the other for baryons. When the elliptic flow is divided by the number of quarks (n$_q$) in each particle (n$_q$ = 2 for mesons and 3 for baryons), and v$_2$/n$_q$ is plotted as a function of the transverse kinetic energy per quark KE$_T$/n$_q$ all particles fall on the same curve as seen in the panel on the right. This constituent quark number scaling establishes that the collective flow developed in the quark stage, rather than a later hadronic stage.
 
After the first few years of RHIC running, PHENIX was able to collect sufficient statistics for its J/psi, heavy-flavor, and photon programs. It observed a suppression of R$_{AA}(J/\psi)$ \cite{PHENIX:2006gsi}, and suppression of nonphotonic electron spectra due to semileptonic decays of heavy-quark hadrons \cite{PHENIX:2005nhb}, while direct-photon yields were observed to scale with the number of nucleon-nucleon collisions. These results further confirmed that the large suppression of high p$_T$ hadrons is a final-state effect due to parton energy loss in the QGP \cite{PHENIX:2012jbv}. Among subsequent publications, PHENIX studied collisions in three small systems with very different intrinsic initial geometries and found that the hydrodynamic model with a QGP was able to describe the three systems \cite{PHENIX:2018lia}, confirming the influence of the initial-state effects on azimuthal correlations.

PHENIX ended its data collection campaign in 2015 and began preparations for its successor, a new fast state-of-the-art detector called sPHENIX. The purpose of sPHENIX is to investigate and understand the microscopic structure of the QGP and to reveal how strongly interacting matter can arise from the interactions of quarks and gluons described by QCD. A description of sPHENIX and predictions for the breadth of its physics program can be found in \cite{Belmont:2023fau}. sPHENIX started collecting data at RHIC in 2023.

\begin{figure}[ht]
\begin{center}
\includegraphics[scale=1.0]{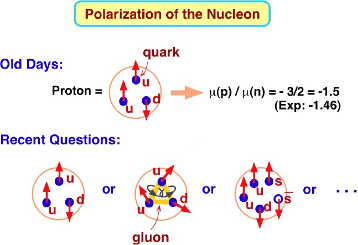}
\end{center}
\caption{The purpose of studies using polarized protons is to understand the origin of the nucleon spin, since only 30\% appears to be carried by quarks. From \cite{Nagamiya:2015raa}.}
\label{figures:Nucleon_polarization}
\end{figure}

With the support of RIKEN, the RHIC spin program has sought to measure the origin of the spin of the proton, which offered additional compelling physics measurements at RHIC for PHENIX and STAR. The data on deep-inelastic lepton scattering by nucleons indicated that quarks carry only about 30$\%$ of the spin of the nucleon \cite{nucleon_spin}. The remaining
$\sim 70\%$ must then come from other mechanisms such as gluon polarization, antiquark polarization, or orbital angular momentum inside the nucleus, as schematically illustrated in Fig.~\ref{figures:Nucleon_polarization}.

Experimentally, the asymmetry between polarized protons is measured in such configurations: p(→) vs. p(←) or p(↑) vs. p(↓). For example, if the gluon carries the nucleon spin, then measurements of the asymmetry of the following channels are of interest: gluon + gluon → jet + jet, and gluon + q
→ high-energy $\gamma$ + $\pi^o$. On the other hand, if anti-quarks carry the nucleon spin, the channels
of interest are: q + $\bar{q}$  → W → e (or $\mu$) + $\nu$ (as in beta-decay).

Once the asymmetry parameters are measured in polarized-proton collisions, the results can be quantitatively connected to the individual components of the spin inside the nucleon. In particular, the PHENIX EM calorimeter was able to measure high-energy photons and e and $\mu$, while STAR measures jet-jet correlations to determine these asymmetry parameters and the various contributions to the spin of the proton. These studies are ongoing at RHIC.

\section{The PHOBOS Experiment at RHIC}
\label{PHOBOS}

After evaluation of the initial proposed experiments, as mentioned earlier, RHIC management decided that there would be two large detectors and two small detectors. In response, a group of physicists from Argonne National Lab, Brookhaven National Lab, INP Krakow Poland, the Massachusetts Institute of Technology, National Central University Chung-Li Taiwan, the University of Illinois at Chicago, the University of Maryland, and the University of Rochester proposed the PHOBOS \footnote{In an earlier round of reviews, the same collaboration proposed a detector called MARS, an acronym for "Modular Array for RHIC Spectrometer”. It was not accepted because of its scope and cost. Theorist John Negele suggested that, when a smaller, less expensive detector was proposed, it should be called PHOBOS, the larger moon of MARS. In this way, if the PHOBOS proposal was still too expensive and an even smaller, less expensive, one had to be built, a name would already be ready for it; DEIMOS, the smaller of the Martian moons!} experiment, as one of the two small detectors. The design was based on the realization that: first, we did not know what would be the most interesting physics; second, shortly after RHIC the much higher energy LHC would begin operations; third, a small group must limit the number of technologies it has to develop; and last and most important, that after the initial studies at RHIC, the small detectors would not be competitive with the two bigger, more expensive ones. 

These considerations led to the following strategy. Build a detector with which one can look at the first physics, particularly the global features, learn as much as possible, and then move to heavy ions at the LHC, using as a guide the results obtained at RHIC. Looking back, it is remarkable how well this strategy worked. In essence, a single technology was used (silicon detectors, with all sensors constructed by one manufacturer in Taiwan and all support structures manufactured from carbon fiber by one group in Poland), and thus crucial parts of PHOBOS were ready at the start of RHIC, so that the experiment obtained publishable results from RHIC within hours of the first collisions \cite{PHOBOS:2000wxz}. It also mapped out the global features of heavy-ion collisions at RHIC energies, and, finally, after the first round of RHIC experiments was completed, many of the members of PHOBOS moved on to continue their studies at the LHC.

\begin{figure}[ht]
\begin{center}
\includegraphics[scale=0.5
]{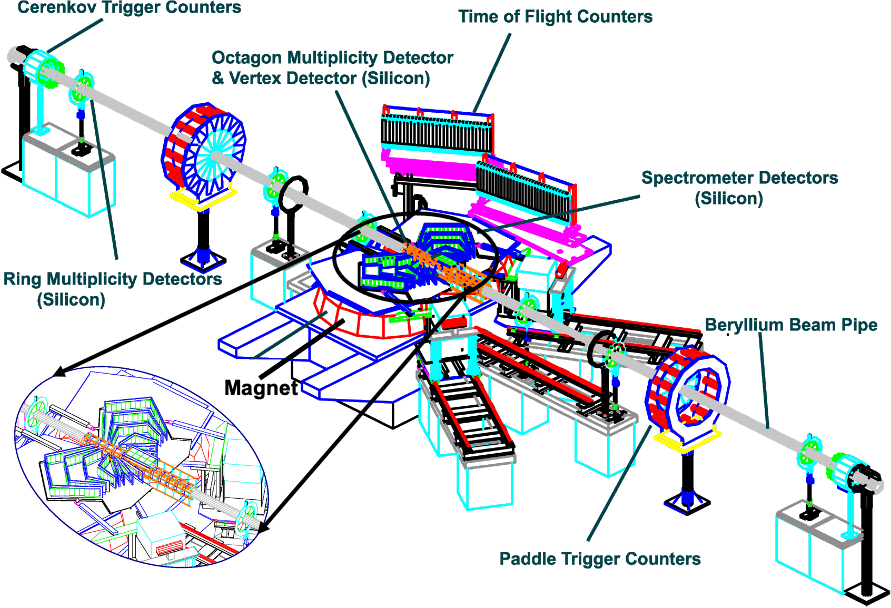}
\end{center}
\caption{The PHOBOS Detector Layout. See text and \cite{PHOBOS_NIM} for reference and details. }
\label{figures:PHOBOS_layout}
\end{figure}

Fig.~\ref{figures:PHOBOS_layout} is a sketch of the PHOBOS detector \cite{PHOBOS_NIM}. Basically, it consists of two systems. One is an array of silicon detectors (labeled Octagon, Vertex and Ring), which detects charged particles produced in almost the entire solid angle around the collision point. Its design was inspired by an analogous experiment studying particle production in p-A collisions carried out in the 1970s at Fermilab \cite{FNAL_Results}. The other system is a small acceptance magnetic spectrometer looking at low-momentum identified particles produced at mid-rapidity, the region where the highest energy density was expected to be produced. 

The idea was that the pseudo-rapidity and azimuthal distributions of all the produced particles would give a global picture of particle production at RHIC, including the energy density produced and the extent to which A-A collisions differ from a superposition of independent pp interactions. On the other hand, the spectrometer was focused on a search for an anomalous production of very low-momentum particles, which would be a sign that a large volume of weakly interacting constituents had been produced. The results from PHOBOS can be found in \cite{PHOBOS:2004zne}.

\begin{figure}[h]
\begin{center}
\includegraphics[scale=0.38]{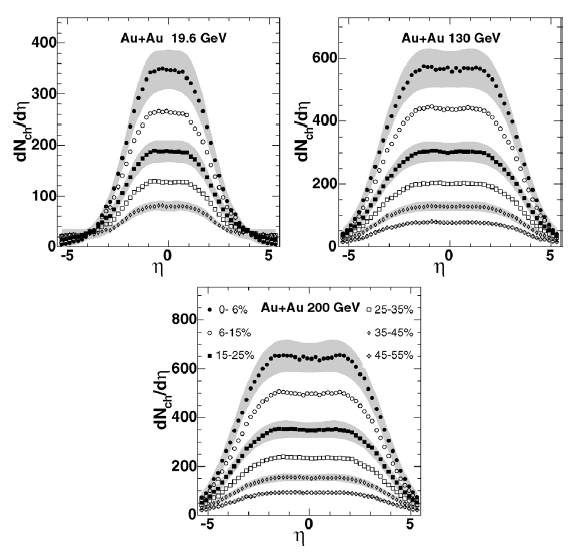}
\end{center}
\caption{Pseudo-rapidity density of charged particles emitted in Au-Au collisions at three different values of the
nucleon–nucleon center-of-mass energy as measured by PHOBOS. Shown are data for bins of centrality as labeled in the bottom panel, binned by the fraction of the total inelastic cross section in each bin. The systematic uncertainties are represented as the gray bands (at the 90$\%$ confidence level). Statistical errors are smaller than
the symbols. From \cite{PHOBOS:2004zne}.}
\label{figures:PHOBOS_dndeta}
\end{figure}

Examples of pseudo-rapidity distributions measured by PHOBOS \cite{PHOBOS:2010eyu} are shown in Fig.~\ref{figures:PHOBOS_dndeta}. The event-by-event knowledge of the density of produced particles in all directions (i.e. both in rapidity and azimuth) encodes the complete particle production process, from the instant of collision, through the evolution of the intermediate state, including the production and decay of a possible QGP, to the hadronization of the finally produced hadrons. Thus, detailed knowledge of such data is crucial for a full understanding of the physics at play in heavy-ion collisions. In the PHOBOS multiplicity array, for example, the measured azimuthal distributions \cite{PHOBOS:2004vcu} showed, together with the other experiments, that there was a liquid intermediate state and that the produced liquid was a strongly interacting fluid with extremely low viscosity-to-entropy ratio, as opposed to the expected weakly interacting one. The measured particle density at mid-rapidity gave information on the dependence on energy and incident system of the maximum energy density released in the collisions.

\begin{figure}[h]
\begin{center}
\includegraphics[scale=0.35]{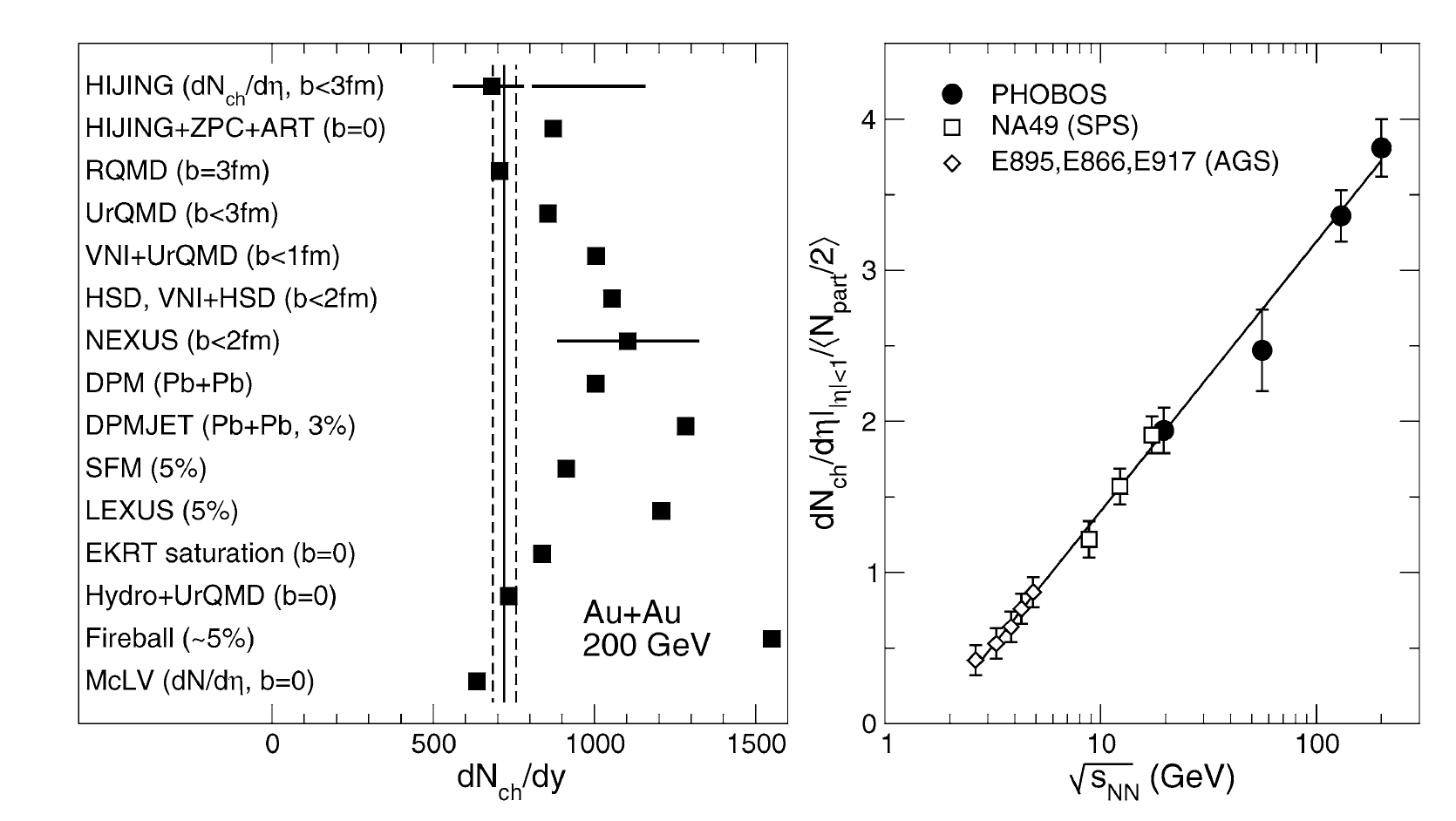}
\end{center}
\caption{Results from PHOBOS on the charged-particle density near mid-rapidity in Au-Au collisions at $\sqrt{s_{NN}}$ = 200 GeV (represented by the vertical line, with the dashed lines denoting the systematic uncertainties), compared with lower energy data and theoretical predictions \cite{Eskola:2001vs}. For details see \cite{PHOBOS:2004zne}.}
\label{figures:PHOBOS_dndy_roots}
\end{figure}

Fig.~\ref{figures:PHOBOS_dndy_roots} compares the particle density measured by PHOBOS with lower energy data and various theoretical model predictions \cite{Eskola:2001vs} made before the onset of RHIC. As can be seen, most models over-predicted the mid-rapidity particle density, and therefore also the produced energy density. However, even the lower values found in the data showed that an energy density high enough to be interesting was being produced, which was a great relief for the community. Other results obtained with the multiplicity array were the discovery of participant scaling of the total particle multiplicity produced in heavy-ion collisions and the observation of extended longitudinal scaling \cite{PHOBOS:2004zne}, which is the extension in rapidity of the phenomenon known as limited fragmentation \cite{Benecke:1969sh} and is direct evidence of saturation effects. Finally, an interesting observation was the fact that overall the data appear to be simpler than the theoretical explanations! For example, the centrality and energy dependences of several observables were found to factorize to a surprising degree \cite{PHOBOS:2010eyu}. Also, no significant discontinuities of any kind as a function of any variable were seen, early signs that if there is a phase transition, it is more likely to be a cross-over rather than a first-order transition.

\begin{figure}[h]
\begin{center}
\includegraphics[scale=0.4]{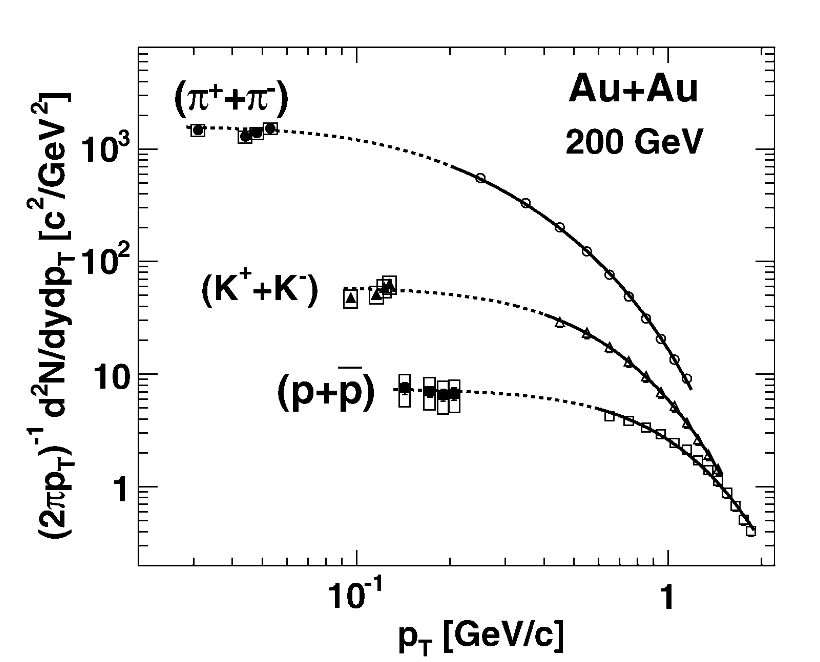}
\end{center}
\caption{Yields of identified charged particles measured by PHOBOS at very low p$_T$,  compared with extrapolations of PHENIX measurements at intermediate p$_T$. Figure is from \cite{PHOBOS:2004zne}. For discussion see text. }
\label{figures:PHOBOS_lowpt}
\end{figure}

The suppression of high transverse momentum particles, called ``jet quenching", was observed using the spectrometer \cite{PHOBOS:2005jpg}. These observations were consistent with measurements made by the other experiments. A unique result obtained with the PHOBOS spectrometer was the
observation that there is no evidence
of an enhancement of low-p$_T$ particles \cite{PHOBOS:2004zne,PHOBOS:2006zpw},  which would be indicative of the presence of unusual long-wave 
phenomena and an early sign that the
medium produced is strongly interacting, 
a very important conclusion that was not fully
appreciated at the time. The observed flattening
of the spectra with p$_T$ increases with the mass
of the particle as seen in Fig.~\ref{figures:PHOBOS_lowpt}, supporting an interpretation of the data in terms of collective
transverse expansion, and is naturally accounted for in the description of the evolution of the system based on hydrodynamic calculations \cite{Kolb:2002ve}.


Although the PHOBOS collaboration, in preparation for the LHC and sPHENIX experiments, ceased to take data after 2005, analysis of the data already collected continued for a number of years. During this period, PHOBOS developed the concept of participant eccentricity and carried out pioneering studies of the long-range nature of the ``ridge" phenomenon.  The former elucidated the importance of the geometric distribution of the nucleons in the nuclei at the instant of collision \cite{PHOBOS:2010ekr}. The latter is evidence of the importance of long-range rapidity correlations of the produced particles \cite{PHOBOS:2009sau}. This eventually led to a unified view of the connection between initial-state fluctuations and final-state correlations in terms of triangularity, triangular flow, and other higher-order components.

\subsection{The BRAHMS Experiment at RHIC}

The smallest of the RHIC experiments was BRAHMS (Broad Range Magnetic Spectrometer) with 40 members and 11 institutions. It was inspired by the CERN ISR experience;
at the time when the ISR experiments were in the planning stage, the most interesting questions were thought to be particle production in the forward and backward directions. Thus, the focus of the ISR detectors was on particles produced in those directions. Right from the start of data taking it became clear that the reality is exactly the opposite and the detectors had to be reconfigured to study particles produced at mid-rapidity! At the time of planning the first round of RHIC experiments, the conventional wisdom was that the most interesting physics will be learned from studies of particles produced at mid-rapidity, those from the hottest region of the produced state. Thus, the PHENIX, PHOBOS, and STAR collaborations all decided to concentrate on the study of these particles. In order that the ISR mistake would not be repeated at RHIC,  the BRAHMS collaboration decided to construct a detector where its greatest asset was its acceptance in the forward and backward directions, where the other detectors are weakest.

\begin{figure}[ht]
\begin{center}
\includegraphics[scale=0.25]{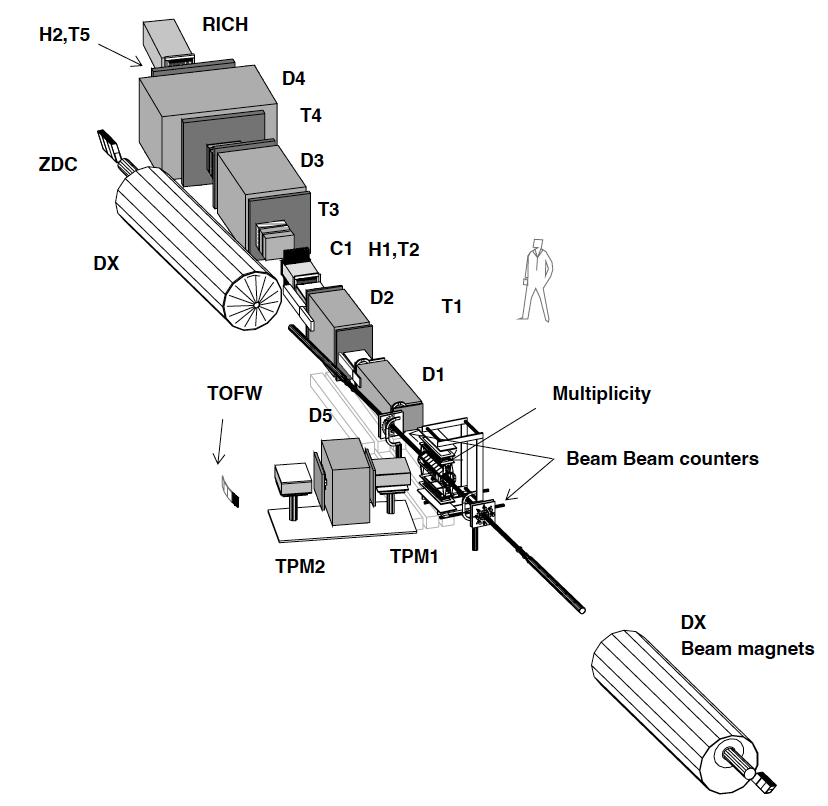}
\end{center}
\caption{BRAHMS detector layout. See \cite{BRAHMS_NIM} for reference and details.}
\label{figures:Brahms_detector_final}
\end{figure}

A sketch of the BRAHMS detector is shown in Fig.~\ref{figures:Brahms_detector_final} with details described in \cite{BRAHMS_NIM}. It consists of two magnetic spectrometers with excellent momentum resolution and hadron identification capability.
Each subtends a small solid angle but can rotate about the collision point and thus can obtain inclusive data for hadron production over a wide range of rapidity.

\begin{figure}[ht]
\begin{center}
\includegraphics[scale=0.6]{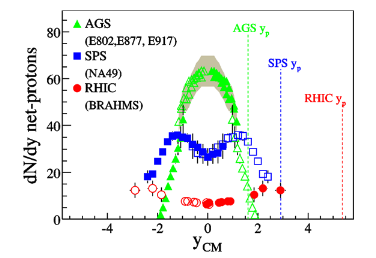}
\end{center}
\caption{Rapidity density of net protons (number of protons minus number of antiprotons) measured at AGS,
SPS, and RHIC for central collisions, from BRAHMS \cite{BRAHMS:2004adc}. BRAHMS was able to extend its unique high-statistics measurements 
to y = 3.5 in $\sqrt{s_{NN}}$ = 200 GeV Au-Au, corresponding to measurements to within 2.3 degrees of the beam direction.}
\label{figures:Brahms_dndyCM}
\end{figure}

With this detector BRAHMS could study, as a function of centrality, identified spectra of p, K, and $\pi$
in the range $0 \leq y \leq 4$ and $0.2 \leq p_T \leq$ 3 GeV/c. 
It should be emphasized that BRAHMS was the only RHIC experiment that could measure, over this large kinematic range, the true rapidity and not just the pseudo-rapidity. The aim was to address questions such as what is the energy available for particle production, what is the longitudinal extent of the near baryon-free medium, what is the medium created at mid-rapidity, how does the "chemistry" of the medium change with rapidity, are small-x effects (saturation) evident in the forward direction. 
The results from BRAHMS can be found in \cite{BRAHMS:2004adc}. An example of BRAHMS results is given in Fig.~\ref{figures:Brahms_dndyCM}, where the measured rapidity density of net protons for central collisions is shown, compared to lower-energy data. From these data it was concluded that at RHIC energies the rapidity loss of the incident protons is approximately 2 units, consistent with expectations from extrapolations of p-A data obtained at Fermilab in the early 1980s.

\section{Pre-RHIC to Post-RHIC}
\begin{figure}[ht]
\begin{center}
\includegraphics[scale=0.4]{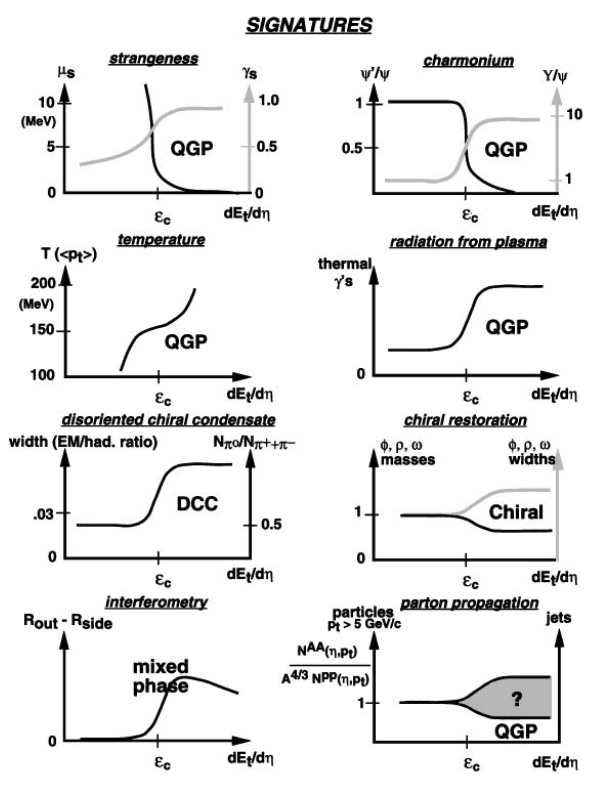}
\end{center}
\caption{Behavior of some of the potential QGP signatures anticipated in 1996, prior to RHIC. From Ref.\cite{Signatures}}
\label{figures:Signatures1996}
\end{figure}

As discussed earlier, in the late 1970s and early 1980s it was thought that in experimentally accessible heavy-ion collisions it should be possible to produce energy densities above 1 GeV/fm$^3$,  and it was expected that at such densities a very interesting phase of QCD should exist: a QCD gas of weakly interacting, free quarks and gluons named the QGP. Furthermore, that there is a first-order transition between a hadron gas and the QGP and therefore it should be easy to discover and study.  Fig.~\ref{figures:Signatures1996} illustrates examples of ``results" that, with luck, would be observed and that would be clear evidence of a phase transition taking place \cite{Signatures}.

It was this kind of thinking that gave rise to the decision to build RHIC and, while waiting, to convert existing proton accelerators to produce as high an energy ion accelerator as possible and use them to collide the beams with stationary nuclei to begin a search for evidence of QGP production.

Reality proved not to be so simple! As discussed in the section of this book describing studies using the AGS accelerator at Brookhaven and the SPS at CERN there were many tantalizing results, but no sign of discontinuities that would show a first-order phase transition, in any variable studied.

By the year 2000, when RHIC became operational and gold nuclei were collided, first at a center of mass energy of 56 GeV/nucleon, and then up to 200 GeV/nucleon, 
the community was divided on their opinion whether there is or is not any evidence for the creation of the QGP  at the AGS or SPS energies.

\begin{figure}[ht]
\begin{center}
\includegraphics[scale=0.45]{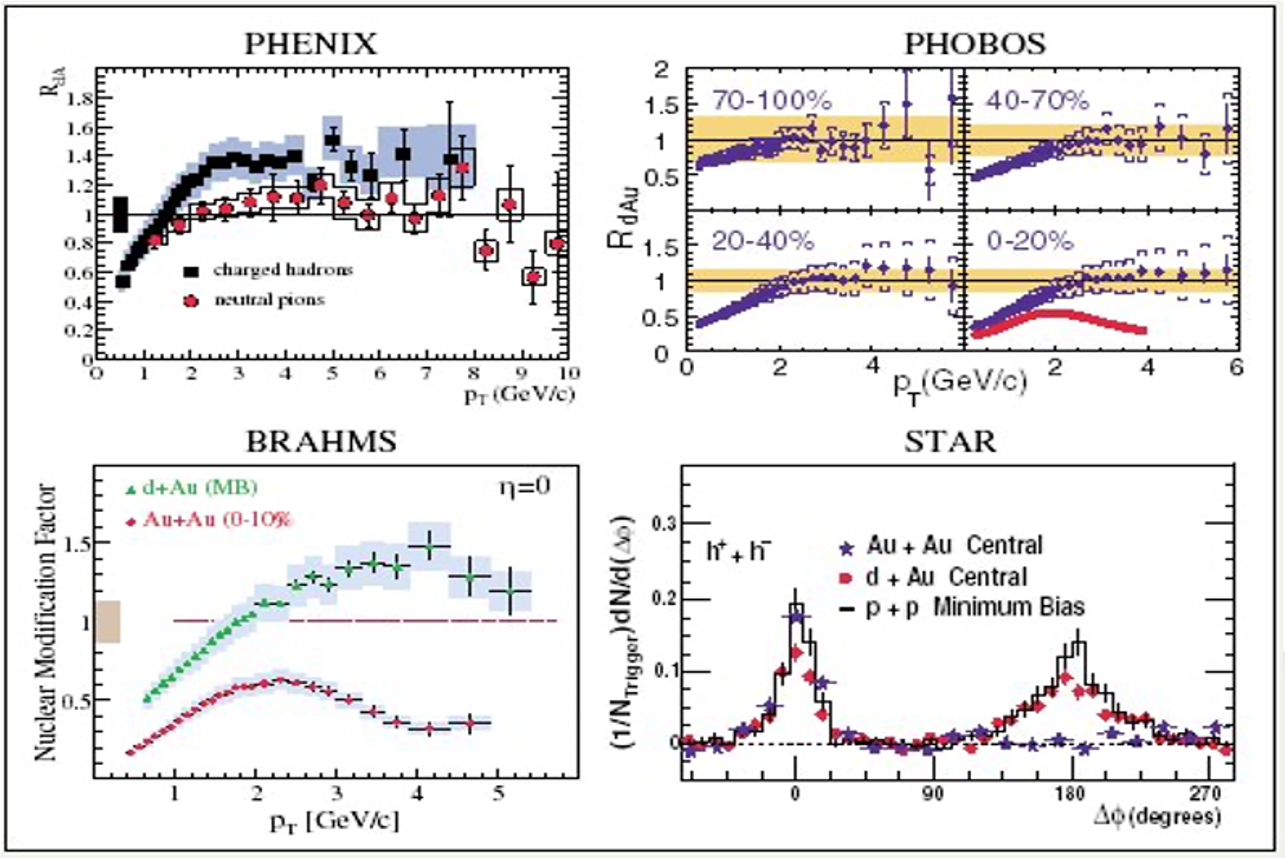}
\end{center}
\caption{
Nuclear modification factor in $\sqrt{s_{NN}}$ =  200 GeV d-Au relative to that from pp collisions for: (upper left) PHENIX inclusive charged hadrons and $\pi^o$s , and (upper right) PHOBOS charged particles for four d-Au centrality bins and in red symbols central (0–6$\%$) Au-Au data. (lower left) BRAHMS charged particles in minimum bias d-Au and central Au-Au at $\sqrt{s_{NN}}$ =  200 GeV. (lower right) STAR angular correlation of charged hadrons in Au-Au, d-Au and pp as labeled. This figure indicates jet quenching in Au-Au but not in d-Au and disappearance of the away-side jet in central Au-Au.  Figure from the cover of Physical Review Letters \cite{PRL_cover}.}
\label{figures:PRL_cover}
\end{figure}

From the first collisions, the RHIC results were a bomb-shell and a game changer! There was no evidence of any discontinuities of observed quantities, which would be indicative of a first-order phase transition, and no signs of any weakly-interacting quark-gluon gas. 
In contrast, the RHIC experiments, to a lesser or greater extent, found that, rather than a weakly-interacting gas being produced, there is strong evidence for the creation of a very strongly-interacting liquid; a liquid with a remarkably low ratio of shear viscosity to entropy. This conclusion followed primarily from two observations; first, the azimuthal angular distributions of the produced particles could be explained by assuming that they obeyed the equations of relativistic hydrodynamics and second, as can be seen in Fig.~\ref{figures:PRL_cover}, that high-energy particles lose a significant fraction of their energy as they traverse the medium produced in the collision, a phenomenon called jet quenching.  At the time this was considered so important that the jet quenching data, from all four RHIC experiments, appeared on the cover of Physical Review Letters (Fig.~\ref{figures:PRL_cover})!  
An important workshop was held at the end of the first round of experiments in May 2004 to discuss the state of knowledge of hot QCD.
The proceedings of the workshop \cite{Rischke:2005ne} and its accompanying "white papers" \cite{White_papers} summarize the results from the four RHIC experiments and provide an excellent overview of the field of relativistic heavy-ion collisions at the beginning of the 21st century.

In retrospect, we have a fairly good understanding of what went wrong with the prediction that the QGP is a weakly interacting gas. There was an erroneous interpretation of some lattice QCD simulations, unrealistic assumptions about how fast the running of the strong coupling constant would lead to a weakly interacting gas, and not understanding well the application of the non-perturbative QCD regime to heavy-ion collisions. Most importantly, we now know that the transition from a hadron gas to the QGP is not a phase transition. It is a cross-over and this explains the smooth behavior of all measurements.

In reality, the fact that a first-order phase transition to a weakly-interacting QGP was not discovered and does not exist is fortuitous! If a non-interacting gas was discovered at RHIC, it would have killed the field. There would be nothing to study!

As it is, RHIC opened a fascinating and rich field of studies that now continues at the LHC.
We still do not thoroughly understand the properties of the QGP, the phenomenon of jet quenching, the physics that takes place in multiparticle production from the instant of collision to the final production of outgoing hadrons, and most importantly, how a complex form of matter, such as the QGP, emerges from simple underlying laws and particles.

\end{document}